# The Effect of Partisanship and Political Advertising
## on Close Family Ties

**Authors:** M. Keith Chen[1*†], Ryne Rohla[2*]

**Affiliations:**

[1]Anderson School of Management, University of California, Los Angeles, CA 90095, USA.

[2]School of Economic Sciences, Washington State University, Pullman, WA 99164, USA.

*Both authors contributed equally to this work.

†Corresponding author. Email: keith.chen@anderson.ucla.edu.

**Abstract:** Research on growing American political polarization and antipathy primarily studies public institutions and political processes, ignoring private effects including strained family ties. Using anonymized smartphone-location data and precinct-level voting, we show that Thanksgiving dinners attended by opposing-party precinct residents were 30-50 minutes shorter than same-party dinners. This decline from a mean of 257 minutes survives extensive spatial and demographic controls. Dinner reductions in 2016 tripled for travelers from media markets with heavy political advertising—an effect not observed in 2015—implying a relationship to election-related behavior. Effects appear asymmetric: while fewer Democratic-precinct residents traveled in 2016 than 2015, political differences shortened Thanksgiving dinners more among Republican-precinct residents. Nationwide, 34 million person-hours of cross-partisan Thanksgiving discourse were lost in 2016 to partisan effects.

**One Sentence Summary:** Smartphone-location data show that mixed-politics families shortened 2016 Thanksgivings, especially in areas with heavy political advertising.

**Introduction:** American political partisanship has risen sharply over the past twenty-five years. More than 55% of Democrats and Republicans described "very unfavorable" feelings toward the opposing party in 2016, up from 17-21% in the mid-1990s; growing numbers of Independents express disfavor with both parties, and rising party defections increase polarization (*1*). Spatial partisan sorting produces increasingly homogeneous electoral "bubbles" at both state and local levels (*2*), and political minorities within these bubbles show reticence to participate in or reveal their party affiliation (*3*).

Animosity toward political rivals is not limited to the ballot box; implicit partisan biases manifest in discriminatory decisions even more frequently than racial or gender biases (*4*). Parents express intolerance of their children dating and marrying across partisan lines (*5*), and observed dating and marital choices segregate more strongly on politics than on physical attributes or personality characteristics (*6*). Political polarization impacts decisions, such as where to work and shop, at rates larger than those caused by race, ethnicity, or religion (*7*).

We study whether politics strain close family ties by measuring family gathering durations. After the historically divisive 2016 presidential election, 39% of American families avoided political conversations during the holidays, an aversion that spanned both party and socioeconomic lines (*8*). We examine Thanksgiving, which, in U.S. election years may bring together family members with differing political views just weeks after votes are cast. Anecdotal evidence suggests that in the wake of the 2016 election, many families canceled or otherwise cut short Thanksgiving plans with their most politically-problematic relatives (*9*).

Several cognitive biases in social and political psychology explain why individuals might limit such interactions. A "partisan selective exposure" motivation occurs when individuals avoid counter-attitudinal political information that might engender cognitive dissonance or harm relationships (*10*). Numerous studies find "belief polarization", where individuals gravitate toward more extreme versions of their own initial positions during discussion of political issues (*11*). Exacerbating this effect, individuals also incorrectly expect others to respond to discussion and debate in the same direction as their own response, anticipating belief convergence rather than polarization (*12*), and attribute a lack of convergence to the bias and irrationality of others while viewing themselves and co-partisans as less ideological than cross-partisans (*13*). Our study examines whether these effects—well-studied in experimental settings among strangers—extend to close family gatherings.

We analyze how political differences affect Thanksgiving through merging two novel datasets. Anonymized smartphone-location data from more than ten million Americans allows observation of actual travel at extremely precise spatial and temporal levels. We combine this with a precinct-level database for the 2016 election to impute presidential voting at the finest spatial resolution possible. By comparing vote shares in an individual's home and Thanksgiving destination precincts, we test the relationship between political disagreement and time expenditure.

To isolate the particular effect of election-year political partisanship from a multitude of demographic and spatial confounds, we construct comparison sets of smartphone users sharing the same home-destination pairs. Our measured effects are neither eliminated nor attenuated by comparing only matched users, suggesting the measured time loss is not an artifact of politically-correlated demographics or spatial sorting. Further, since political advertising polarizes opinions (*14*) and heightens dislike for opposing parties (*15*), we compare partisan rifts between comparable users who fall just on opposite sides of media-market boundaries. Accounting for political advertising more than triples our measured Thanksgiving effect in 2016, but not in 2015 before ads were run. This non-effect of yet-to-be-run ads acts as a political placebo test, further bolstering the argument that our measured Thanksgiving losses stem from political partisanship rather than pre-existing demographic or spatial confounds.

**Data Collection and Validation:** We collect precinct-level results for the 2016 presidential election through internet scraping and by contacting Secretaries of State, Boards of Election, and individual County Clerks via email, phone, fax, or in person. Finally, we match vote totals to precinct polygonal shapefiles using Geographic Information Systems (GIS) software. The resulting dataset covers approximately 172,000 precincts across 99.9% of counties nationally (Fig. 1A).

Political advertising data are from Kantar Media's Campaign Media Analysis Group (*16*), and count every U.S. presidential television ad aired in all 210 Nielsen Designated Market Areas after June 12, 2016, including ads purchased directly by campaigns or by outside groups such as Political Action Committees. Data from the 2010 Decennial Census and the Census Bureau's 2012-2015 American Community Survey form the basis of demographic controls.

Location data relies on numerous smartphone apps, aggregated by SafeGraph, a company which builds and maintains anonymized geospatial data sets for more than 10 million U.S. smartphones. These data consist of "pings", each identifying the coordinates of a particular smartphone at a moment in time. Our primary analysis includes 21 billion pings from November 2016, and 4.5 billion from November 2015.

To merge datasets, we infer the precinct and census block of each smartphone user's "home", based on that user's pings between 1am and 4am over the three weeks before Thanksgiving. This procedure identifies more than 6 million approximate home locations in November 2016 (Fig. 1B), which we then link with corresponding precinct two-party vote shares and census demographics. Similarly, a user's Thanksgiving location is based on their modal location between 1pm and 5pm (November 24[th], 2016 and 26[th], 2015).

By construction, this sample is representative of the 77% of Americans who own smartphones, raising the question of whether our sample is politically representative of the American electorate as a whole. We test this by assigning to each resident a vote ratio proportional to the 2016 two-party vote share of their home precinct. A resident of a precinct that recorded 150 Clinton and 50 Trump votes, for example, would be assigned 0.75 Clinton and 0.25 Trump votes. Fig. 1(C) compares these votes against actual 2016 two-party vote shares for each state. The 45-degree line represents where states would lie if the SafeGraph sample politically matched the distribution of American voters. Our imputed votes are accurate to within 1 percentage-point in 33 states and within 5 percentage-points in all states. Nationally, the data suggest a two-party Democratic vote share of 51.6%, compared to the actual share of 51.1%.

**Empirical Analysis:** We first examine whether, conditional on traveling for Thanksgiving dinner, the partisan distance between a home and destination affect that dinner's duration. We restrict our sample to residents who were home both in the morning and the night of Thanksgiving, but who traveled for Thanksgiving dinner, to focus our analysis on travelers who could control the duration of their visits. In Table 1 we estimate the following equation:

$$duration_{ij} = \alpha + \beta \; mismatch_{ij} + \gamma \; F_{ij} + \varepsilon_{ij}$$

where

$$mismatch_{ij} = P_i \, (1 - P_j) + (1 - P_i) \, P_j$$

In this specification, $duration_{ij}$ is the minutes traveler $i$ spent with host $j$ on Thanksgiving, and $F_{ij}$ is a set of fixed-effects that form groups of people defined by pairs of home ($i$) and destination ($j$) locations. $P_i$ and $P_j$ are the two-party vote shares associated with home precincts for $i$ and $j$, where $P_i = (dem_i \, / \, (dem_i + rep_i))$. Using $P_i$ and $P_j$, $mismatch_{ij}$ is the imputed probability that persons $i$ and $j$ voted for different candidates in 2016. In all tables, regressions control for progressively finer ($i,j$) location pairs culminating in 5-digit geohash boxes, a global grid of rectangular areas approximately three miles per side.

To control for confounds including demographics, distance, and travel time, our regressions compare Thanksgiving durations between travelers with the same home and destination areas. For example, regression 3 compares two travelers if and only if they both live in zipcode $X$ and visit zipcode $Y$. The coefficient of interest, $\beta$, measures the reduction in Thanksgiving durations between travelers within the same $F_{ij}$ comparison groups, but who likely voted differently than their Thanksgiving hosts. Standard errors are clustered at the home-precinct × destination-precinct level. We use progressively tighter spatial controls to control for both demographics and travel distance simultaneously.

Results in Table 1 indicate that families that were likely to have voted for different presidential candidates spent approximately 30 to 50 fewer minutes together—off an average Thanksgiving dinner of 4.2 hours—after controlling for both travel distance and location-correlated demographics. As we add finer spatial controls, our estimate of $\beta$ remains fairly stable, with a point estimate of 56.3 +/- 14.6 minutes under our tightest geohash-5 controls. In online Table A1, we report qualitatively identical results when demographics such as race, age, education, income, and employment are controlled for separately.

We examine the two components of $mismatch_{ij}$, $P_i \, (1 - P_j)$ and $(1 - P_i) \, P_j$, to separately measure the effect of voting disagreement among Democratic-precinct residents (DPRs) visiting Republican-precinct residents (RPRs) and vice versa. Table 2 demonstrates that, conditional on traveling, DPRs shorten their visits to RPR hosts by approximately 20 to 40 minutes, while RPRs shortened their visits to DPRs by approximately 50 to 70 minutes. $F$-test results indicate that these estimates are statistically different (p < 0.0001 in 4 of 5 specifications), with RPRs shortening their cross-party stays by more than DPRs.

Investigating whether these effects interact with local political advertising, we find that cross-partisan Thanksgiving dinners are further shortened by around 2.6 minutes on average, for every thousand political advertisements aired in the traveler's home media market (Table 3). Some media markets in swing states saw more than 26,000 ads over the course of the campaign, implying an approximately 69-minute shorter Thanksgiving for vote-mismatched families in Orlando, for example, compared to those in markets without advertising. While this effect may not be solely due to advertising, which may be correlated with other campaign activities such as rallies, campaign visits, and fundraising efforts, these results bolster the conclusion that measured effects on Thanksgiving dinner duration likely stem from an increased intensity and salience of partisan differences.

Table S1 supports this finding by performing a placebo test concerning whether advertisements in 2016 affected Thanksgiving dinner behavior the year before airing. Regardless of whether we pool smartphone users or split the sample into DPRs and RPRs, we find no evidence of pre-existing partisan effects in regions that witnessed high advertising levels. While our empirical results estimate briefer Thanksgiving dinners among cross-partisan gatherings in both years, the ad-related amplification of this effect is present only in 2016—both in statistical significance and magnitude—supporting our conjecture that the main effect is most likely political in nature.

Examining destination choices, our data suggest that travelers did not change plans to reduce political divisions from 2015 to 2016. Among travelers who traveled in both years—the strongest possible control for demographic and spatial confounds—we observe no appreciable difference in the distribution of likely political mismatch (Fig. S1). This finding suggests that travelers were more likely to change the duration of Thanksgiving gatherings than to change the destination.

Finally, Tables S2 and S3 estimate linear probability models for the choice of whether to travel for Thanksgiving, in both 2015 and 2016. When comparing matched residents living within 1.5 miles of each other, DPRs reduced their likelihood of travel between 2015 and 2016 by 2 percentage-points more than comparable RPRs, an effect which increases significantly in areas with heavy political advertising.

Examining only those residents included in both the 2015 and 2016 dataset yields qualitatively similar results. Among residents at home on Thanksgiving morning in both years, 56.4% traveled for Thanksgiving in 2015, while 51.9% traveled in 2016 (n=28,890, Fisher's $p < 0.0005$). Accompanying this difference is a reduction in Thanksgiving duration for those cross-partisan dinners that still occurred. Comparing travelers who went to the same location both years (n=1,271), we estimate that politically-mismatched gatherings declined by 42.1 +/- 41.4 minutes. While this small sample size precludes statistically significance, this estimate is very close to our findings in Table 1.

Aggregating across the 77% of American adults who own smartphones (*17*), our results suggest that partisan differences cost Americans 73.6 million person-hours of Thanksgiving time in 2016, 47.8% from DPRs and 52.2% from RPRs. Political advertising-related partisanship comprised 15.9 million of lost person-hours, 46.3% from DPRs and 53.7% from RPRs. Altogether, an estimated 33.9 million person-hours of cross-partisan discourse were eliminated, perhaps creating a feedback mechanism by which partisan segregation reduces opportunities for close cross-party conversations.

Our findings have several implications, both to the literature and for campaign policy. Following the 2016 election, anecdotal media reports and online social media behavior (*18*) demonstrated an avoidance of personal confrontations over political issues among Democratic voters, findings our study corroborates. RPRs, however, were more sensitive to partisan differences at Thanksgiving dinners, an effect that supports findings of greater partisan-selective exposure among Republicans in news media consumption (*19*). Our results suggest that partisan polarization extends in quantitatively meaningful ways to close family settings, and that political advertising and related campaign efforts can exacerbate these fissures. As abbreviated Thanksgiving gatherings tend to accumulate in regions with greater campaign activity, policies designed to shorten campaigns may reduce the private costs of political polarization.

**Acknowledgments:** The authors thank seminar participants at UCLA, Washington State, Northwestern, Simon Fraser, and Stanford Universities, as well as Elisa Long for helpful comments, and A. Hoffman, R. Squire and N. Yonack at Safegraph for data access and technical assistance. **Funding:** There are no funding sources related to this study. **Author contributions:** M.K.C and R.R. designed and implemented the study, acquired the data, and drafted and revised the manuscript.  **Competing interests:** Neither author has any competing interests. **Data and Materials Availability:**  All data and code are available at: http://www.anderson.ucla.edu/faculty/keith.chen/datafilm.htm

**Author contributions:** M.K.C. and R.R. designed and implemented the study, and drafted and revised the manuscript.

**List of Supplementary Materials:**

Figure S1

Tables S1-S3

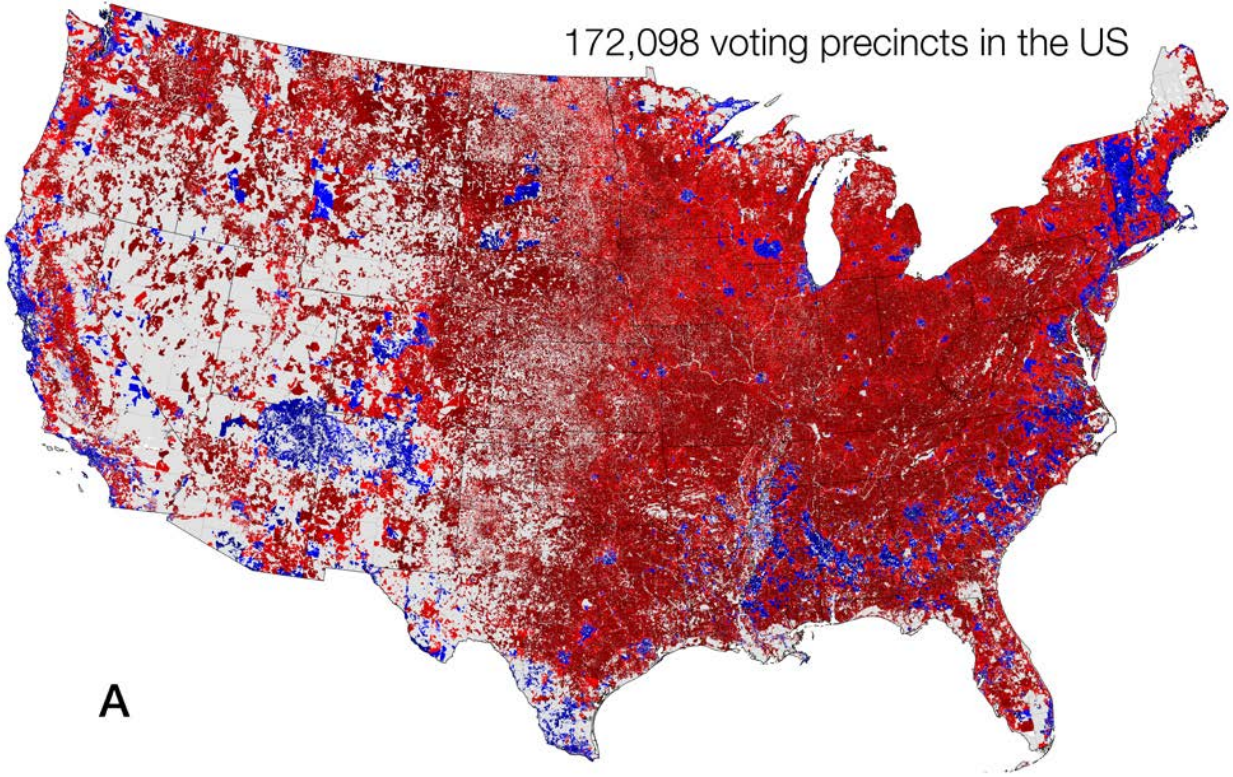

172,098 voting precincts in the US

A

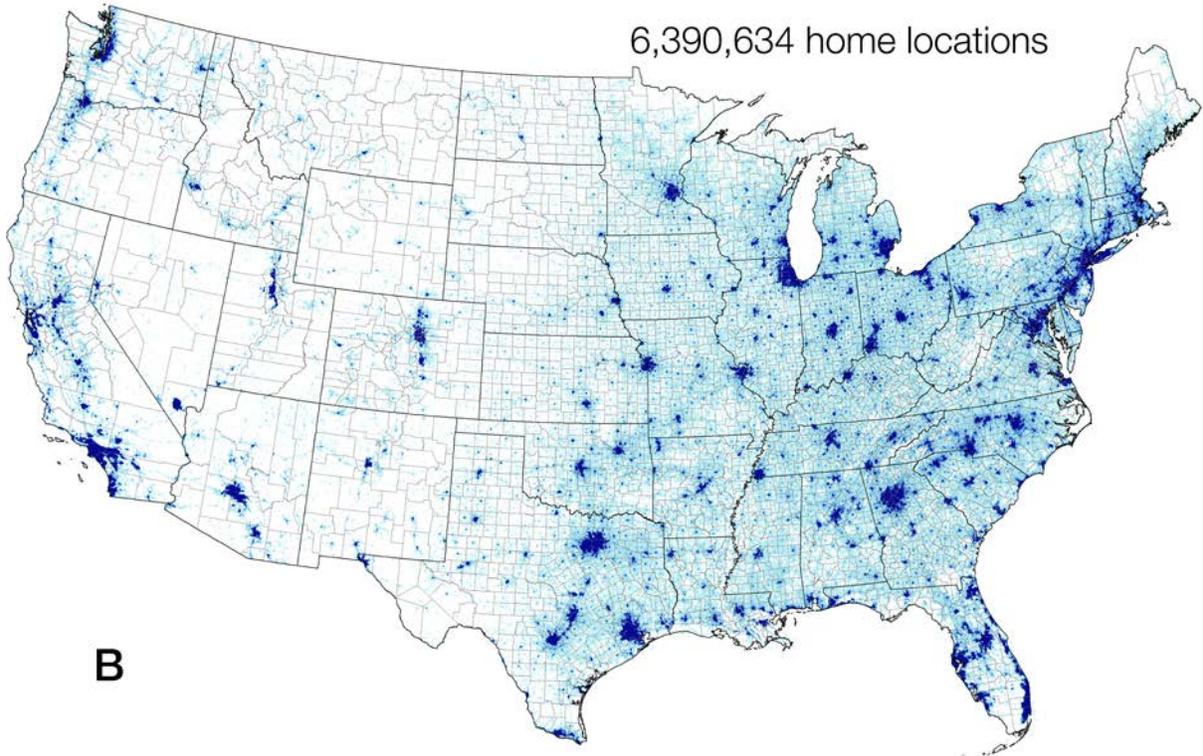

6,390,634 home locations

B

**Fig. 1. Sampling and imputation validation.**
(**A**) Results of the 2016 U.S. presidential election by precinct (excludes unpopulated census blocks). (**B**) 2016 home locations of smartphone users in the sample. (**C**) Correlation between actual two-party vote share by state (x-axis) and vote share predicted (y-axis) using each smartphone user's home-precinct. Nationally this predicts a 0.516 Clinton vote share, compared to an actual vote share of 0.511. Highlighted are the two most Democratic-leaning, Republican-leaning, and largest prediction-error states.

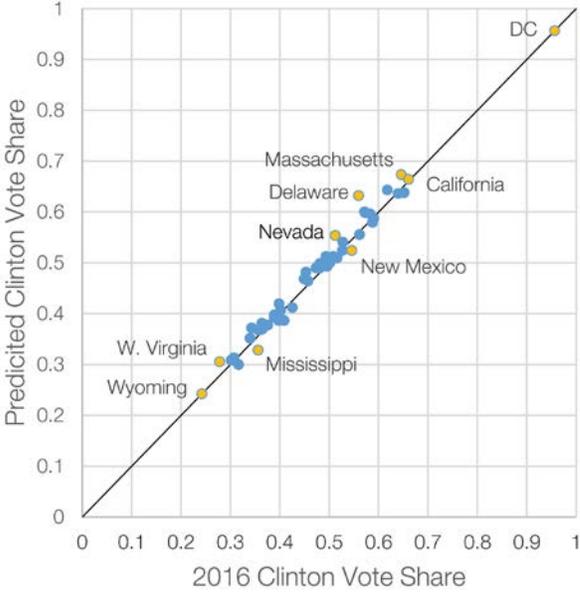

C

| Variable: | 1<br>Thanksgiving<br>Duration (min) | 2<br>Thanksgiving<br>Duration (min) | 3<br>Thanksgiving<br>Duration (min) | 4<br>Thanksgiving<br>Duration (min) |
|---|---|---|---|---|
| Prob. of political<br>mismatch | -21.58***<br>(2.226) | -38.04***<br>(2.952) | -45.23**<br>(8.696) | -56.26**<br>(14.55) |
| | | | | |
| Observations: | 642,962 | 642,962 | 642,962 | 642,962 |
| R-squared: | 0.0003 | 0.0660 | 0.458 | 0.661 |
| Fixed-effects: | none | county pairs | zip-code pairs | geohash-5 pairs |
| Num. of FE groups: | | 35,507 | 302,716 | 414,950 |

Table 1. Effect of political mismatch on Thanksgiving duration. Each column estimates the effect of voting disagreement on 2016 Thanksgiving dinner duration. All results use linear regressions with fixed effects controlling for an individual's home location × Thanksgiving destination. Regressions 1 through 4 control for progressively finer pairs, culminating in a 5-digit geohash, a square grid of approximately 3 miles per side. The mean duration of Thanksgiving dinner was 257 minutes [SD: 162 minutes]. The average probability of voting mismatch was 0.44 [SD: 0.10]. Standard errors are reported in parentheses and clustered at the precinct x precinct level. Significance levels: *** $p<0.001$, ** $p<0.01$, * $p<0.05$.

| Variable: | 1<br>Thanksgiving<br>Duration (min) | 2<br>Thanksgiving<br>Duration (min) | 3<br>Thanksgiving<br>Duration (min) | 4<br>Thanksgiving<br>Duration (min) |
|---|---|---|---|---|
| Prob. D → R<br>trvlr → host | -5.60*<br>(2.454) | -23.44***<br>(3.207) | -30.16**<br>(9.358) | -44.53**<br>(15.73) |
| Prob. R → D<br>trvlr → host | -38.74***<br>(2.555) | -53.47***<br>(3.314) | -60.23***<br>(9.455) | -69.20***<br>(16.14) |
| | | | | |
| F (R→D ≠ D→R) | 0.0001 | 0.0001 | 0.0001 | 0.0572 |
| Observations: | 642,962 | 642,962 | 642,962 | 642,962 |
| R-squared: | 0.0003 | 0.0662 | 0.458 | 0.661 |
| Fixed-effects: | none | county pairs | zip-code pairs | geohash-5 pairs |
| Num. of FE groups: | | 35,507 | 302,716 | 414,950 |

Table 2. Asymmetric effects of political mismatch. Each column estimates the effect of voting disagreement between travelers and hosts (DPR to RPR and RPR to DPR) on 2016 Thanksgiving dinner duration. The F-test p-value tests for equality between coefficients, to test for an asymmetric mismatch effect. The average probability of DPRs eating at a RPR-hosted dinner, and vice-versa, was 0.221 and 0.215, respectively [both SD = 0.10]. Standard errors are reported in parentheses and clustered at the precinct x precinct level. Significance levels: *** $p<0.001$, ** $p<0.01$, * $p<0.05$.

| Variable: | 1 Thanksgiving Duration (min) | 2 Thanksgiving Duration (min) | 3 Thanksgiving Duration (min) | 4 Thanksgiving Duration (min) |
|---|---|---|---|---|
| Prob. of political mismatch | -21.58*** (2.226) | -14.40*** (2.588) | | |
| Prob. D → R trvlr → host | | | -5.604* (2.454) | 4.117 (2.879) |
| Prob. R → D trvlr → host | | | -38.74*** (2.555) | -33.68*** (2.978) |
| Political ads (1K ads / mrkt) | | 1.334*** (0.185) | | 1.349*** (0.185) |
| Prob. Pol. Mis. × Pol. ads | | -2.645*** (0.393) | | |
| Prob. D → R × Pol. ads | | | | -3.237*** (0.417) |
| Prob. R → D × Pol. ads | | | | -2.122*** (0.439) |
| Observations: | 642,962 | 642,962 | 642,962 | 642,962 |
| R-squared: | 0.0003 | 0.0004 | 0.0003 | 0.0004 |

Table 3. Political advertising heightens partisan effects. Each column estimates the effect of voting disagreement between travelers and hosts (DPR to RPR and RPR to DPR) on 2016 Thanksgiving dinner duration. Columns 2 and 4 explore whether political advertising heightens these effects. Media markets in swing states like Florida saw more than 26,000 ads in 2016, Standard errors are reported in parentheses and clustered at the precinct x precinct level. Significance levels: *** p<0.001, ** p<0.01, * p<0.05.

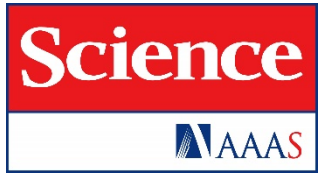

Supplementary Materials for

**The Effect of Partisanship and Political Advertising
on Close Family Ties**

**Authors:** M. Keith Chen[1][*][†], Ryne Rohla[2][*]

**Affiliations:**

[1]Anderson School of Management, University of California, Los Angeles, CA 90095, USA.

[2]School of Economic Sciences, Washington State University, Pullman, WA 99164, USA.

*Both authors contributed equally to this work.

†Corresponding author. Email: keith.chen@anderson.ucla.edu.

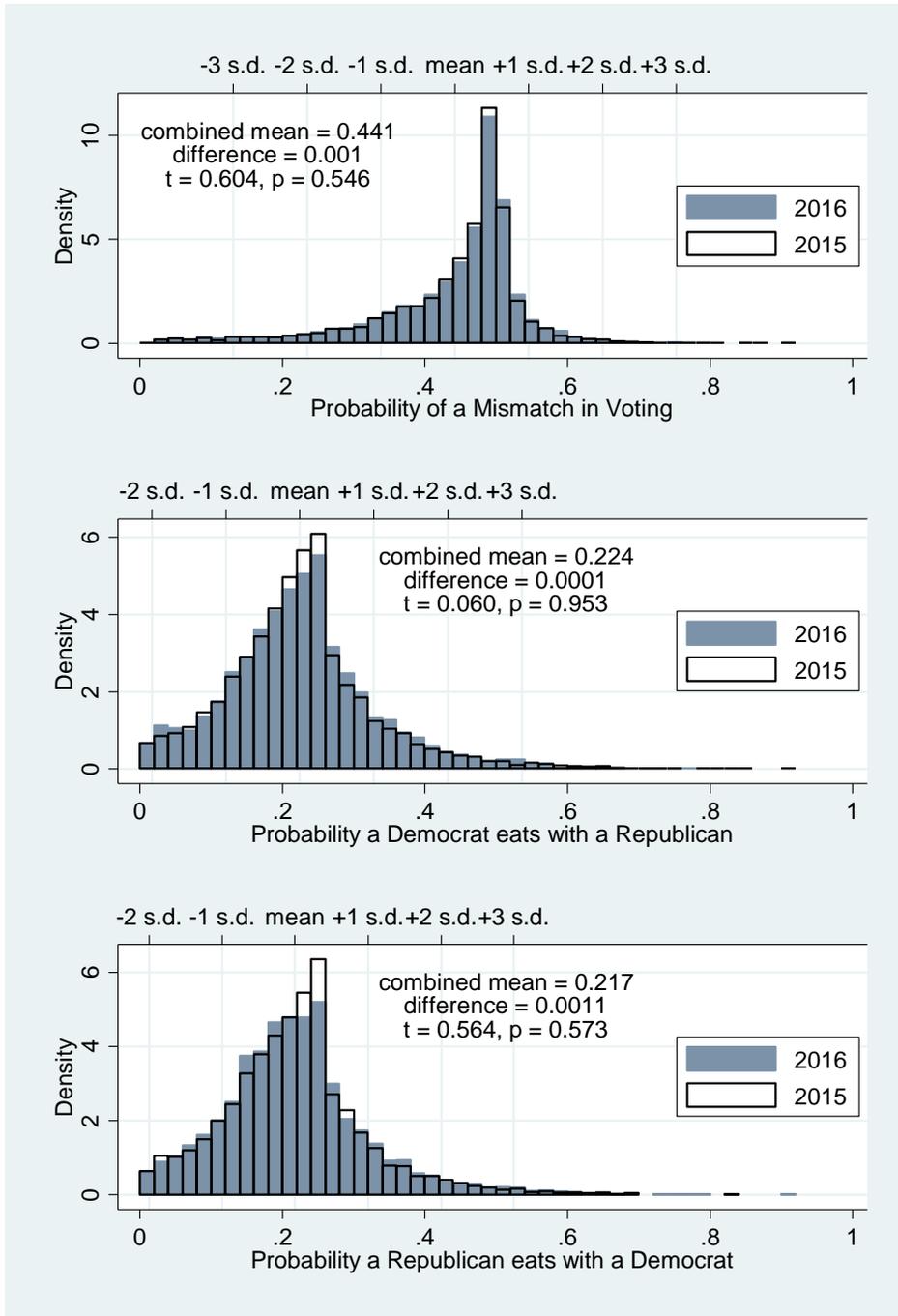

**Fig. S1. The (non)effect of partisanship on Thanksgiving destination choice.** The distributions of voting mismatch for Thanksgivings 2015 and 2016 for people who traveled in both years. (**A**) The distribution of the probability that a person voted differently from their Thanksgiving hosts, for both 2015 and 2016. (**B,C**); the two ways mismatch can occur; a DPR traveler eating with a RPR host (**B**), or vice versa (**C**). T-tests confirm that conditional on traveling for Thanksgiving dinner, the partisan difference of travelers and hosts did not change significantly between 2015 and 2016.

| Variable: | 1 Thanksgiving Duration (2016) | 2 Thanksgiving Duration (2015) | 3 Thanksgiving Duration (2016) | 4 Thanksgiving Duration (2015) |
|---|---|---|---|---|
| Prob. of political mismatch | -14.40*** | -23.88*** | | |
| | (2.588) | (6.208) | | |
| Prob. D → R trvlr → host | | | -4.117* | -14.34* |
| | | | (2.879) | (7.036) |
| Prob. R → D trvlr → host | | | -33.68*** | -35.68*** |
| | | | (2.978) | (7.338) |
| Political ads (1K ads / mrkt) | 1.334*** | 0.543 | 1.349*** | 0.523 |
| | (0.185) | (0.407) | (0.185) | (0.408) |
| Prob. Pol. Mis. × Pol. ads | -2.645*** | -0.402 | | |
| | (0.393) | (0.876) | | |
| Prob. D → R × Pol. ads | | | -3.237*** | -0.900 |
| | | | (0.417) | (0.939) |
| Prob. R → D × Pol. ads | | | -2.122*** | 0.222 |
| | | | (0.439) | (1.025) |
| | | | | |
| Observations: | 642,962 | 56,201 | 642,962 | 56,201 |
| R-squared: | 0.0004 | 0.0008 | 0.0004 | 0.0010 |

Table S1. The (non)effect of advertising on Thanksgiving in 2015. Each column estimates the effect of voting disagreement on 2016 Thanksgiving dinner duration, and the interaction of that effect with political ads run in that media market in 2016. While the sample of tracked smartphones is smaller in 2015, columns 2 and 4 show effects of political mismatch in 2015 which are well estimated and quantitatively similar to 2016 in both symmetric and asymmetric specifications. Importantly though, this effect interacts with political advertising only in 2016, and not in 2015 (before those ads were run). This suggests the main effect is driven by political differences, and not unobservable differences between more or less mismatched families. The mean duration of Thanksgiving dinner in 2015 was 196 minutes, and the average probability of opposite-voting political mismatch was 0.42 with a SD of 0.12. Standard errors are clustered at the precinct x precinct level and reported in parentheses, with significance levels: *** p<0.001, ** p<0.01, * p<0.05.

| Variable: | 1 Thanksgiving not at home | 2 Thanksgiving not at home | 3 Thanksgiving not at home | 4 Thanksgiving not at home |
|---|---|---|---|---|
| Post election | -0.143*** | -0.139*** | -0.134*** | -0.127*** |
| (year = 2016) | (0.00357) | (0.00375) | (0.00379) | (0.00401) |
| Home 2-party | -0.0811*** | -0.0587*** | -0.0335*** | -0.0436*** |
| vote (Dem) | (0.00582) | (0.00641) | (0.00726) | (0.00792) |
| PE × H2PV | -0.0034 | -0.0073*** | -0.0128* | -0.0223*** |
| | (0.00602) | (0.00631) | (0.00637) | (0.00668) |
| | | | | |
| Observations: | 2,163,307 | 2,163,307 | 2,163,307 | 2,163,307 |
| R-squared: | 0.006 | 0.0139 | 0.0321 | 0.0810 |
| Fixed-Effects: | none | county | zip code | 5-digit geohash |
| Num. of Groups: | | 3,100 | 30,245 | 117,405 |

**Table S2. Partisanship and Thanksgiving travel: 2015 & 2016.** Each column estimates the effect of political leanings on the choice to eat Thanksgiving dinner at home or away, and how this effect differs between the pre-election 2015 and post-election 2016 Thanksgivings. All regressions are fixed-effect linear probability regressions, where fixed effects control for the location of a person's home. Standard errors are clustered at the home-precinct level and reported in parentheses, with significance levels: *** $p<0.001$, ** $p<0.01$, * $p<0.05$.

| Variable: | 1<br>Thanksgiving<br>not at home | 2<br>Thanksgiving<br>not at home | 3<br>Thanksgiving<br>not at home | 4<br>Thanksgiving<br>not at home |
|---|---|---|---|---|
| Post election | -0.153*** | -0.149*** | -0.144*** | -0.136*** |
| (year = 2016) | (0.00420) | (0.00440) | (0.00446) | (0.00473) |
| Home 2-party | -0.0878*** | -0.0662*** | -0.0373*** | -0.0476*** |
| vote (Dem) | (0.00686) | (0.00753) | (0.00854) | (0.00937) |
| PE × H2PV | 0.0120 | 0.0070 | 0.0008 | -0.0100 |
| | (0.00708) | (0.00741) | (0.00750) | (0.00787) |
| Political ads | -0.00147** | 0.00434 | -0.000757 | -0.00005 |
| (1K ads / mrkt) | (0.000533) | (0.00436) | (0.00105) | (0.00139) |
| PE × H2PV | -0.00418*** | -0.00369*** | -0.00348*** | -0.00304** |
| × Pol. ads | (0.000967) | (0.00103) | (0.00104) | (0.00108) |
| H2PV | 0.00165 | 0.00194 | 0.00110 | 0.00107 |
| × Pol. ads | (0.000931) | (0.00102) | (0.00113) | (0.00123) |
| PE × Pol. ads | 0.00269*** | 0.00244*** | 0.00234*** | 0.00206*** |
| | (0.000550) | (0.000586) | (0.000589) | (0.000750) |
| | | | | |
| Observations: | 2,162,992 | 2,162,992 | 2,162,992 | 2,162,992 |
| R-squared: | 0.006 | 0.0139 | 0.0321 | 0.0810 |
| Fixed-Effects: | none | county | zip code | 5-digit geohash |
| Num. of Groups: | | 3,099 | 30,244 | 117,400 |

Table S3. Political advertising and Thanksgiving travel: 2015 & 2016. Each column estimates the effect of political leanings on the choice to eat Thanksgiving dinner at home or away, and how this effect differs between the pre-election 2015 and post-election 2016 Thanksgivings, and between areas that saw more or less political advertising in 2016. All regressions are fixed-effect linear probability regressions, where fixed effects control for the location of a person's home. For example, Regression 3 can be interpreted as saying that in 2016 all residents reduced their Thanksgiving travel propensity, DPRs reduced Thanksgiving travel more than RPRs living in their same zipcode, and this effect was more pronounced for DPRs residing in areas with high quantities of political advertising. Standard errors are clustered at the home-precinct level and reported in parentheses, with significance levels: *** $p<0.001$, ** $p<0.01$, * $p<0.05$.

**Online Appendix Materials:**

| Variable: | 1 Thanksgiving Duration (min) | 2 Thanksgiving Duration (min) | 3 Thanksgiving Duration (min) | 4 Thanksgiving Duration (min) |
|---|---|---|---|---|
| Prob. of political mismatch | -17.02*** (2.331) | -27.46*** (3.054) | -41.98** (8.716) | -43.81** (14.98) |
| White proportion (Census block) | -20.03*** (6.059) | -3.793 (6.741) | -12.06 (12.92) | -14.79 (23.75) |
| Black proportion (Census block) | -16.26** (6.197) | 9.108 (6.928) | 5.087 (13.36) | -6.264 (24.47) |
| Hisp. proportion (Census block) | -14.04* (6.229) | -4.813 (6.945) | -9.534 (13.32) | -18.86 (24.31) |
| Asian proportion (Census block) | -10.82 (6.505) | -8.341 (7.201) | -10.70 (14.07) | -15.35 (24.81) |
| Foreign-born (Census tract) | 37.57*** (2.728) | -15.89*** (3.705) | -23.11* (10.10) | -17.79 (14.18) |
| Male proportion (Census bl. gp.) | 16.34*** (4.202) | 12.40* (4.438) | 7.148 (8.673) | 7.110 (13.97) |
| Median age (Census tract) | 0.347*** (0.037) | -0.049 (0.045) | -0.104 (0.104) | -0.066 (0.174) |
| Urban proportion (Census tract) | -4.224*** (0.965) | -0.286 (1.444) | 3.709 (4.893) | -3.352 (16.98) |
| Rural proportion (Census tract) | -11.93*** (1.198) | -4.548** (1.477) | -1.194 (4.065) | -2.138 (14.74) |
| Median HH Inc. (Census tract, $1K) | -0.030*** (0.009) | -0.127*** (0.012) | -0.162*** (0.032) | -0.236*** (0.052) |
| Unemployment (Census bl. gp.) | 83.40*** (6.377) | 17.09* (6.927) | 19.99 (13.63) | 18.21 (21.49) |
| Avg. Commute Time (Census bl. gp., min) | 0.052 (0.034) | -0.075 (0.047) | -0.123 (0.111) | -0.310 (0.184) |
| | | | | |
| Observations: | 642,358 | 642,358 | 642,358 | 642,358 |
| R-squared: | 0.0003 | 0.0666 | 0.458 | 0.661 |
| Fixed-effects: | none | county pairs | zip-code pairs | geohash-5 pairs |
| Num. of groups: | | 35,446 | 302,371 | 414,518 |

**Table A1. Effect of political mismatch on Thanksgiving duration in 2016 with demographic controls.** Each column is an estimate of the effect of voting disagreement on the length of Thanksgiving dinner in 2016. All regressions are fixed-effect linear regressions, where fixed effects control for the pair of locations where an individual lives and ate Thanksgiving dinner. Regressions running from left to right control for progressively finer pairs of areas, culminating in 5-digit geohash boxes, a grid of boxes roughly 3 miles per side. The number of comparison groups these fixed-effects entail is listed for each regression. The mean duration of Thanksgiving dinner was 257 minutes, and the average probability of opposite-voting political mismatch was 0.44 with a SD of 0.10. Standard errors are clustered at the precinct x precinct level and reported in parentheses, with significance levels: *** $p<0.001$, ** $p<0.01$, * $p<0.05$.

**Supplementary Map:**

## The Effect of Partisanship and Political Advertising on Close Family Ties
Minutes of Thankgiving Dinner Time Lost to Political Partisanship by County, 2016

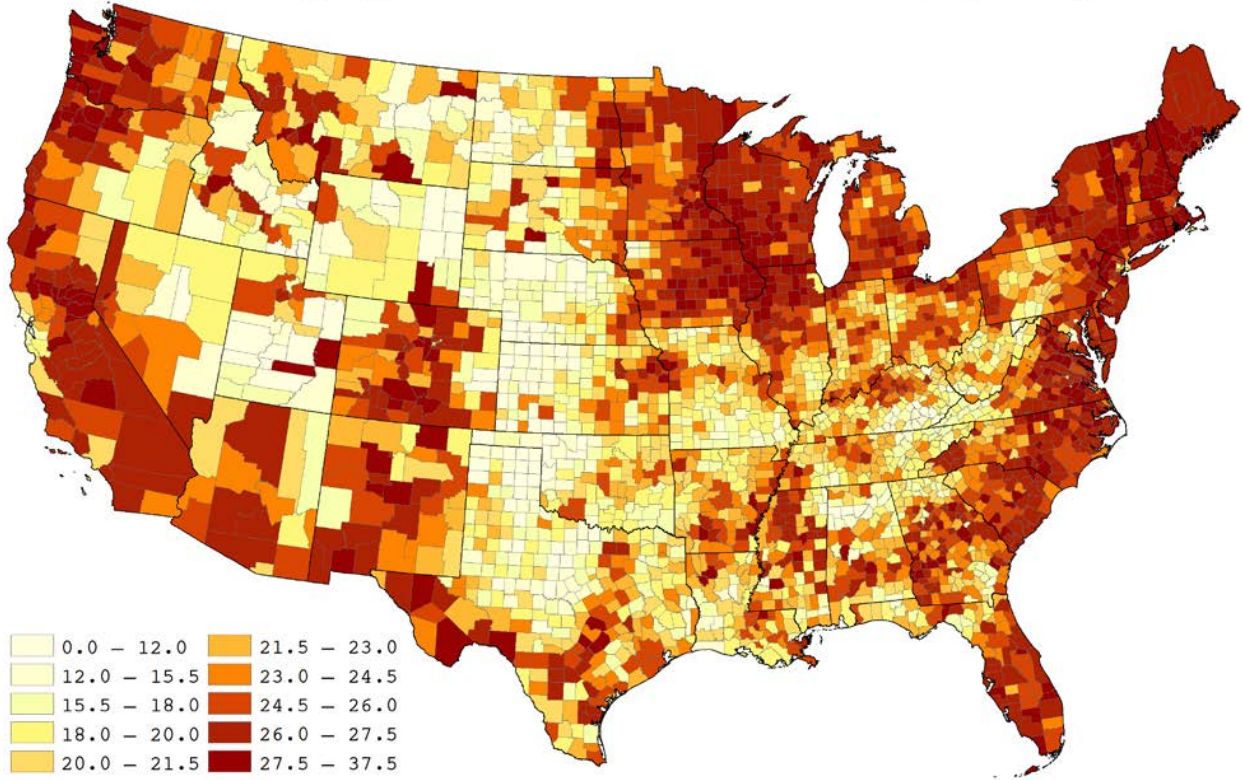

| | |
|---|---|
| 0.0 — 12.0 | 21.5 — 23.0 |
| 12.0 — 15.5 | 23.0 — 24.5 |
| 15.5 — 18.0 | 24.5 — 26.0 |
| 18.0 — 20.0 | 26.0 — 27.5 |
| 20.0 — 21.5 | 27.5 — 37.5 |

*Notes: Darker color = more Thanksgiving minutes lost due to partisanship-shortened gatherings*
*Estimates obtained from anonymized smartphone location data*